\documentclass[epj]{webofc}
\usepackage{slashed}     
\usepackage[utf8]{inputenc}
\usepackage[varg]{txfonts}   
\usepackage{booktabs}
\usepackage{xcolor}
\definecolor{darkred}{rgb}{0.4,0.0,0.0}
\definecolor{darkgreen}{rgb}{0.0,0.4,0.0}
\definecolor{darkblue}{rgb}{0.0,0.0,0.4}
\usepackage[bookmarks,linktocpage,colorlinks,
    linkcolor = darkred,
    urlcolor  = darkblue,
    citecolor = darkgreen]{hyperref}

\wocname{EPJ Web of Conferences}
\woctitle{Lattice2017}
\newcommand{\matrixel}[3]{\left< #1 \vphantom{#2#3} \right| #2 \left| #3 \vphantom{#1#2} \right>}
\newcommand{\abs}[1]{\left| #1 \right|}
\newcommand{\avg}[1]{\left< #1 \right>}
\newcommand{\eq}[1]{\begin{equation} #1 \end{equation}}

\newcommand{\mc}[1]{\mathcal{#1}} 
\begin{document}
%
\selectlanguage{english}
\title{%
  Heavy Domain Wall Fermions:
  The RBC and UKQCD charm physics program
}
\author{%
\firstname{Peter A} \lastname{Boyle}\inst{1} \and
\firstname{Luigi} \lastname{Del Debbio}\inst{1} \and
\firstname{Andreas} \lastname{J{\"u}ttner}\inst{2} \and
\firstname{Ava} \lastname{Khamseh}\inst{1} \and
\firstname{Justus Tobias} \lastname{Tsang}\inst{1}\fnsep\thanks{Speaker, \email{J.T.Tsang@ed.ac.uk}} \and
\firstname{Oliver} \lastname{Witzel}\inst{1,3} 
}
\institute{%
Higgs Centre for Theoretical Physics, School of Physics \& Astronomy, University of Edinburgh, Edinburgh, EH9 3FD, United Kingdom
\and
School of Physics and Astronomy, University of Southampton, Southampton, SO17 1BJ, United Kingdom
\and
Department of Physics, University of Colorado, Boulder, CO 80309, USA
}
\abstract{%
We review the domain wall charm physics program of the RBC and UKQCD collaborations based on simulations including ensembles with physical pion mass. We summarise our current set-up and present a status update on the decay constants $f_D$, $f_{D_s}$, the charm quark mass, heavy-light and heavy-strange bag parameters and the ratio $\xi$. 
}
\maketitle

\section{Introduction}\label{intro}
In this work we summarise RBC/UKQCD's domain wall charm physics project with $N_f=2+1$ flavours of dynamical domain wall fermions. The goal of our ongoing efforts is to provide calculations of non-perturbative quantities which allow, combined with experimental input, for the extraction of CKM~\cite{Kobayashi:1973fv} matrix elements. In addition we aim to extract the charm quark mass $m_c$. Finally we are working towards extending the reach in the heavy quark mass beyond $m_c$ to allow for extrapolations to $b$-physics.

In this status report we focus on decay constants, neutral meson mixing and the determination of the charm quark mass. In the following we will first introduce these observables, before detailing our set-up (see Section \ref{sec:ens}) and presenting preliminary results (see Section \ref{sec:res}). We conclude with an outlook of future steps.

\subsection{Observables} \label{subsec:obs}
The decay constant $f_{D}$ ($f_{D_s}$) of the $D$ ($D_s$) meson is defined by
\eq{
  \matrixel{0}{A_{cq}^\mu}{D_q(p)} = f_{D_q} p^\mu_{D_q},
  \label{eq:decayconst}
}
where $q=d,s$ and the axial vector current is defined as $A_{cq}^\mu = \overline{c}\gamma_\mu \gamma_5 q$.  
The experimentally measurable decay widths $\Gamma\left(D_{(s)} \to l\nu_l\right)$ can be combined with these decay constants, allowing to extract the CKM matrix elements $\abs{V_{cq}}$ for $q=d,s$. More precisely, we can write
\eq{
  \Gamma\left(D_{(s)} \to l\nu_l\right) = \abs{V_{cq}}^2 f^2_{D_{(s)}} \mc{K} + \mc{O}(\alpha_{EM}),
}
where $\mc{K}$ are some known kinematical expressions. So, neglecting electromagnetic effects, the decay width factorises.

Neutral mesons are known to mix and their mass difference can be related to the CKM matrix via
\eq{ \label{eqn:deltampheno} 
  \Delta m_q=\frac{G_F^2m_W^2}{6 \pi^2}|V_{tq}^*V_{tb}|^2 S_0(x_t)\,\eta_{2B} \,f_{P_q}^2\, m_{P_q}\, B_{P_q}
}
The Inami-Lim function $S_0(x_t)$ \cite{Inami:1980fz} and $\eta_{2B}$~\cite{Buras:1990fn} are known functions in perturbation theory, so by combining knowledge of the non-perturbative \emph{bag parameter} $B_{P_q}$ with the experimental value of $\Delta m_q$ ($q=d,s$) we can determine $\abs{V_{tq}^*V_{tb}}$. The bare bag parameter $B^\mathrm{bare}_{P_q}$ and the four-quark operator $O_{VV+AA}$ are defined as
\eq{
  \label{eqn:bagparam}
  B^\mathrm{bare}_P=\frac{\matrixel{\bar{P}_q^0}{O_{VV+AA}}{P_q^0}}{8/3 f_{P_q}^2m_{P_q}^2} = \frac{\matrixel{\bar{P}_q^0}{\left(\bar{b}\gamma^\mu(1-\gamma_5)q\right)\left(\bar{b}\gamma^\mu(1-\gamma_5)q\right)}{P_q^0}}{8/3 f_{P_q}^2m_{P_q}^2}.
}

When we consider neutral $B_{(s)}$ meson mixing, and noting that our action is chirally symmetric, we can construct a renormalisation independent ratio given by $\xi$, 
\eq{
  \xi=\frac{f_{B_s}\sqrt{B^\mathrm{bare}_{B_s}}}{f_{B}\sqrt{B^\mathrm{bare}_{B}}}.
}
When combined with the experimental measurements, $\xi$ allows to extract the ratio of CKM matrix elements $|V_{td}/V_{ts}|$ which enters as an important constraint in fits of the unitarity triangle.

The final observable we aim to calculate in this report is the charm quark mass $m_c$. To achieve this, the bare charm-like quark masses are first non-perturbatively renormalised. For this we employ a variant of the Rome-Southampton scheme with non-exceptional kinematics and two choices of projectors (i.e. the $(\slashed q, \slashed q)$ and the $(\gamma_\mu,\gamma_\mu)$ schemes~\cite{Martinelli:1994ty,Sturm:2009kb,Aoki:2010pe}. In this scheme, the charm quark mass $m^\mathrm{RI/SMOM}_c(a)$ is then set for each value of the lattice spacing $a$ by requiring that it reproduces the physical value of some hadronic quantity such as $m_{D_{(s)}}$ or $m_\mathrm{\eta_c}$~\cite{Patrignani:2016xqp}. We then take the continuum limit ($a\to 0 $) to remove lattice artifacts. Finally, the result is then matched to some continuum scheme, such as $\overline{m_c}(\mu)$, i.e. determined in the $\overline{MS}$ scheme at some scale $\mu$. We emphasise that different choices in the details of the renormalisation scheme, as well as choosing different hadronic quantities to set the quark mass, allow for different approaches to the continuum which must reproduce the same result.

\section{Ensemble and measurement parameters}\label{sec:ens}

\begin{table}
  \caption{Main parameters of the used ensembles. All ensembles were generated with the Iwasaki gauge action with $2+1$ flavours of domain wall fermions in the sea.}
  \begin{center}
    \begin{tabular}{|cc|cccc|}
      \hline
      $\beta$ & Name & $L^3 \times T /a^4$ &  $a^{-1}[\mathrm{GeV}]$ & $m_\pi[\mathrm{MeV}$] & $m_\pi L $ \\\hline
      2.13 & C0  & $48^3 \times 96$    & 1.7295(38) & {\bf 139}  & 3.9\\
      2.13 & C1  & $24^3 \times 64$    & 1.7848(50) & 340  & 4.6 \\
      2.13 & C2  & $24^3 \times 64$    & 1.7848(50) & 430  & 5.8 \\\hline
      2.25 & M0  & $64^3 \times 128$   & 2.3586(70) & {\bf 139}  & 3.8\\
      2.25 & M1  & $32^3 \times 64$    & 2.3833(86) & 300  & 4.1 \\
      2.25 & M2  & $32^3 \times 64$    & 2.3833(86) & 360  & 4.9 \\
      2.25 & M3  & $32^3 \times 64$    & 2.3833(86) & 410  & 5.5 \\\hline
      2.31 & F1  & $48^3 \times 96$    & 2.774(10)$\hphantom{0}$  & 235  & 4.1 \\
      \hline 
    \end{tabular}
  \end{center}
  \label{tab:ens}
\end{table}

\begin{table}
  \caption{Light quark parameters of the used ensembles. The valence light quark mass was always chosen to be the sea light quark mass. The strange sea quark mass was partially quenched to its physical value where it was known before the run was started. The domain wall parameters $L_s$ and $M_5$ were also chosen to be the same for valence and sea quarks.}
  \begin{center}
    \begin{tabular}{|c||l|ll||cc|}
      \hline
      masses & \multicolumn{1}{c|}{light} & \multicolumn{2}{c||}{strange} & \multicolumn{2}{c|}{simulated strange quark mass}\\
      Name & $am_l^\mathrm{sea} = am_l^\mathrm{val}$ &  $am_s^\mathrm{sea}$ & $am_s^\mathrm{phys}$ & Run 1 & Run 2 \\\hline
      C0  & 0.00078 & 0.0362 & 0.03580(16)  & sea & phys\\
      C1  & 0.005 & 0.04 & 0.03224(18)  & sea \& phys & phys\\
      C2  & 0.01 & 0.04 & 0.03224(18)  & phys & phys\\\hline
      M0  & 0.000678 & 0.02661 & 0.02540(17) & sea & phys\\
      M1  & 0.004 & 0.03 & 0.02477(18) &  sea \& phys & phys\\
      M2  & 0.006 & 0.03 & 0.02477(18) &  phys & phys \\
      M3  & 0.008 & 0.03 & 0.02477(18) & - & phys \\\hline
      F1  & 0.002144 & 0.02144 & 0.02132(17) & sea & phys \\
      \hline 
    \end{tabular}
  \end{center}
  \label{tab:light}
\end{table}

The ensembles present in this study~\cite{Allton:2007hx,Allton:2008pn,Blum:2014tka,Boyle:2017jwu} use the Iwasaki gauge action~\cite{Iwasaki:2011np} and the domain wall fermion action~\cite{Kaplan:1992bt,Shamir:1993zy} with either the Shamir kernel~\cite{Kaplan:1992bt,Shamir:1993zy} (C1-2, M1-3) or the M\"obius kernel~\cite{Brower:2004xi} (C0, M0, F1). The basic parameters of the ensembles are presented in Table \ref{tab:ens}. The properties of the light quark sector (light and strange) are given in Table \ref{tab:light}.  The heavy propagators are generated using the M{\"o}bius kernel in all cases.

We have two different sets of simulation data on the same ensembles, which in the following will be referred to as \emph{Run 1} and \emph{Run 2}. The main difference between \emph{Run 1} and \emph{Run 2} is the choice of domain wall parameters for the heavy propagators. We determined a set of domain wall parameters ($M_5=1.6$ with bound $am_c \lesssim 0.4$) that are suitable for the simulation of charm quarks in refs \cite{Cho:2015ffa,Boyle:2016imm} and presented a first large scale simulation result for this choice in ref \cite{Boyle:2017jwu}. In parallel we investigated the effect of stout smearing~\cite{Morningstar:2003gk} the gauge fields when generating the heavy propagators. We presented this study and preliminary results in refs \cite{Tsang:2016iky,Boyle:2015kyy}. The exact choices of the heavy masses and the measurement statistics are given in table \ref{tab:heavy}. Note that $N_\mathrm{conf}$ indicates the number of statistically independent configurations whilst $N_\mathrm{hits}$ gives the number of measurements per configuration. These different measurements on the same configurations are obtained by placing sources on every $(T/a)/N_\mathrm{hits}$ time planes and are then averaged into one effective measurement per configuration.
For \emph{Run 1} we used $Z_2$ wall-sources~\cite{Boyle:2008rh} for all propagators, whilst in \emph{Run 2} we used Gaussian smeared $Z_2$ wall sources to achieve earlier overlap with the ground state.

We emphasise that in all heavy-light and heavy-strange quantities discussed, we have a mixed action since the discretisation for the light and strange propagators differs from that of the heavy propagators. This is taken into account by carrying out the mixed action non-perturbative renormalisation. This has however not been completed yet for all cases. So instead we will only present bare quantities or build ratios where the renormalisation constants cancel. However, in the case of \emph{Run 1}, the effect of the mixed action is expected to be very mild, as it only originates from a change in the domain wall parameter choice~\cite{Boyle:2016imm,Boyle:2017jwu}.

\begin{table}
  \caption{Heavy quark parameters of the used ensembles and statistics. The columns $N_\mathrm{conf}$ and $N_\mathrm{hits}$ give the number of de-correlated configurations and number of measurements per configuration respectively.}
  \begin{center}
    \begin{tabular}{|c|ccr||ccr|}
      \hline
      Name & \multicolumn{3}{c||}{Run 1 ($M_5=1.6$, no stout smearing)} & \multicolumn{3}{c|}{Run 2 ($M_5=1.0$, stout smearing)}\\
      & $am_h$ & $N_\mathrm{conf}$ & $N_\mathrm{hits}$ & $am_h$ & $N_\mathrm{conf}$ & $N_\mathrm{hits}$ \\\hline
      C0  & 0.30, 0.35, 0.40 &  88 & 48  & 0.51, 0.57, 0.63, 0.69 &  90 & 48\\
      C1  & 0.30, 0.35, 0.40 & 100 & 32  & 0.50, 0.58, 0.64, 0.69 & 100 & 32\\
      C2  & 0.30, 0.35, 0.40 & 101 & 32  & 0.51, 0.59, 0.64, 0.68 & 101 & 32\\\hline
      M0  & 0.22, 0.28, 0.34, 0.40 & 80  & 32 & 0.41, 0.50, 0.59, 0.68 & 82 & 64 \\
      M1  & 0.22, 0.28, 0.34, 0.40 & 83  & 32 & 0.41, 0.50, 0.59, 0.68 & 83 & 32 \\
      M2  & 0.22, 0.28, 0.34, 0.40 & 76  & 16 & 0.41, 0.50, 0.59, 0.68 & 76 & 32 \\
      M3  & \multicolumn{3}{c||}{-}      &  0.41, 0.50, 0.59, 0.68 & 81 & 32\\\hline
      F1  & 0.18, 0.23, 0.38, 0.33, 0.40 & 82 & 48 & 0.32, 0.41, 0.50, 0.59, 0.68 & 98 & 48 \\
      \hline 
    \end{tabular}
  \end{center}
  \label{tab:heavy}
\end{table}

\section{Results}\label{sec:res}
Since this is a status report, all the presented data are still work in progress and are to be taken as an indication of the status of the calculation.
\subsection{Decay constants}\label{subsec:decayconst}
\begin{figure}
  \includegraphics[width=\textwidth]{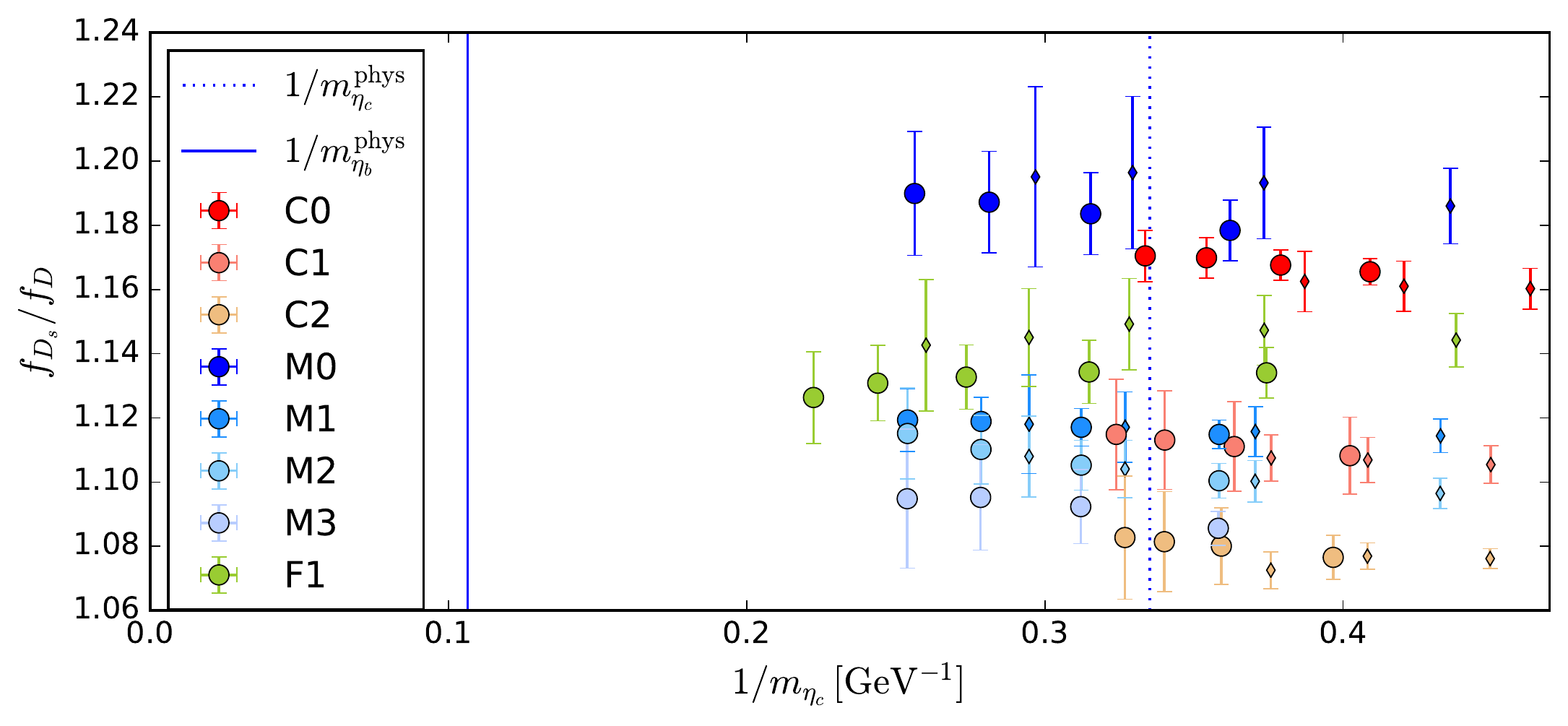}
  \caption{Preliminary results for the ratio of heavy-strange and heavy-light decay constants as a function of the inverse $\eta_c$ mass. The small diamonds, show data from the old ``unsmeared'' runs, the large circles of the new ``smeared'' runs.}
\label{fig:fDsfD}
\end{figure}

Figure \ref{fig:fDsfD} shows our results for the ratio of decay constants, obtained from uncorrelated double exponential fits. The smaller diamonds (Run 1) show the data presented in ref \cite{Boyle:2017jwu}; the larger circles the new data from Run 2. We observe that we can now reach the physical charm quark mass (dotted blue line) also on the coarse ensembles and can reach further in the heavy quark mass, potentially allowing for an extrapolation to the physical $b$-quark mass (solid blue line) in the future. We also notice that the statistical error on the new results has improved as a combination of Gaussian source/sink smearing and the stout smearing of the charm-like propagators. In addition, we have doubled the statistics on the ensemble M0 resulting in a further improvement in the statistical error. Furthermore we note that in agreement with our previous result, we see a very mild behaviour as a function of the inverse $\eta_c$ mass.

\subsection{Bag and $\xi$ parameters}\label{subsec:bag}
\begin{figure}
  \centering
  \includegraphics[width=.45\textwidth]{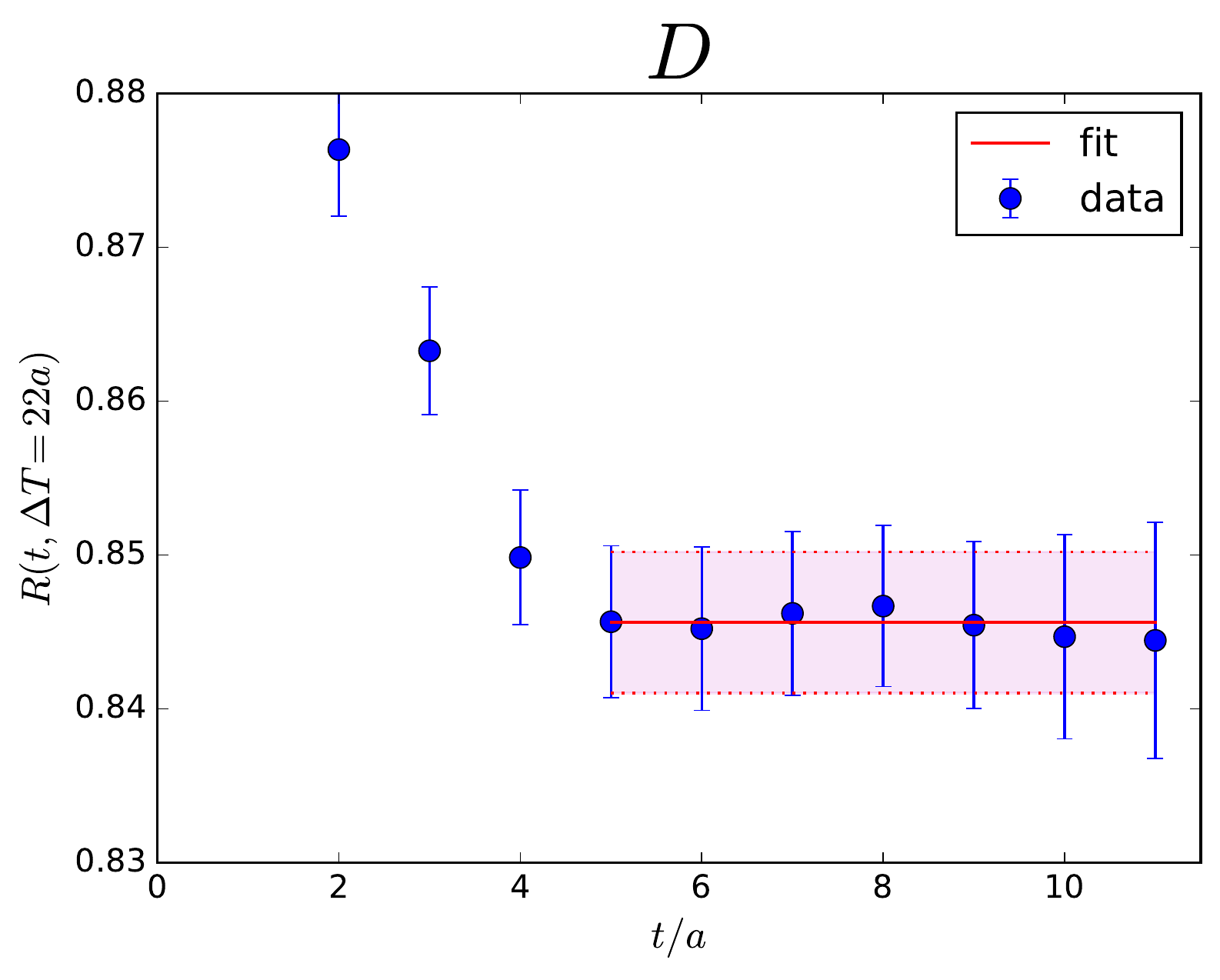}
  \includegraphics[width=.45\textwidth]{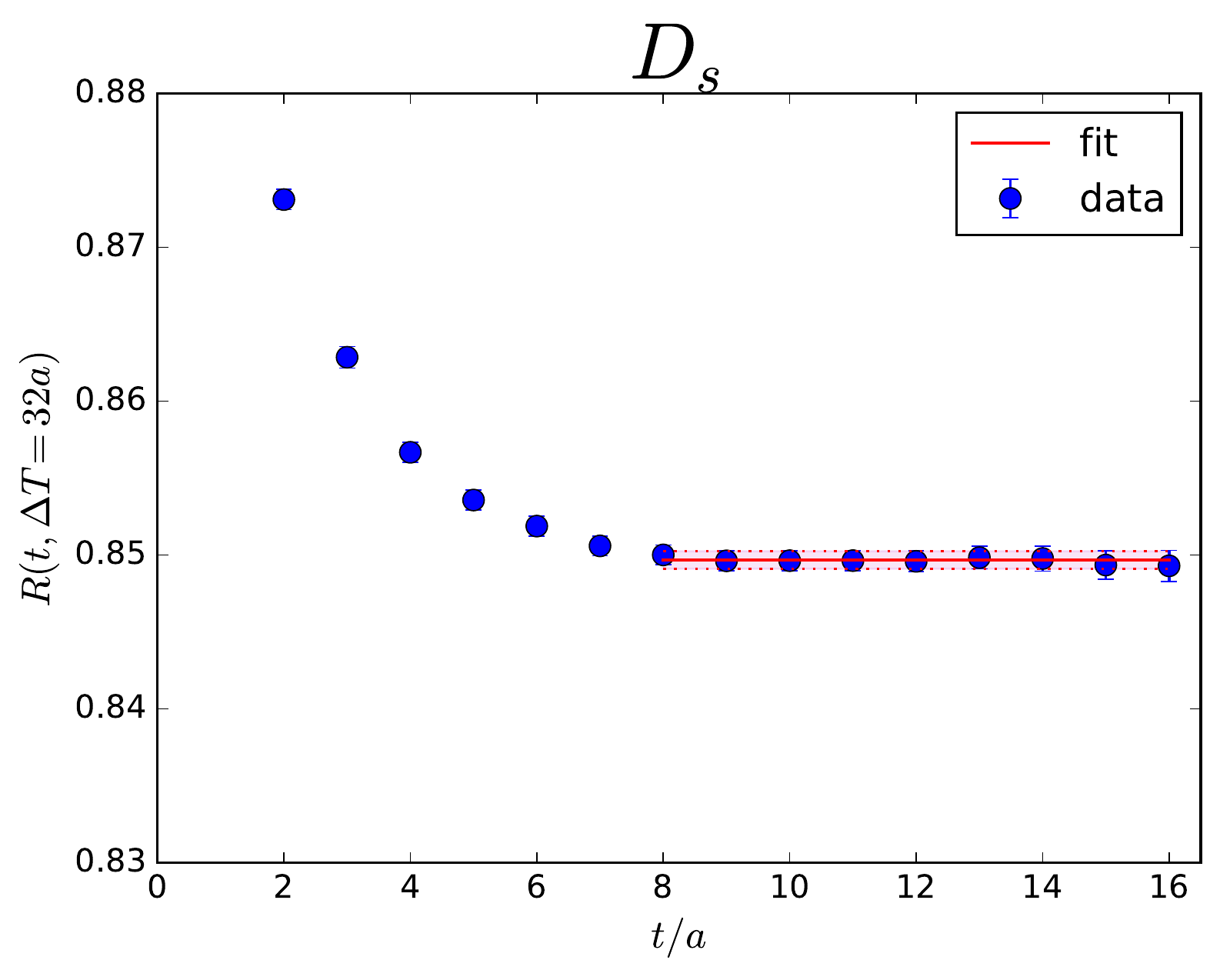}

  \caption{Example fit of the bag parameter for the heaviest mass point on the M0 ensemble. The left (right) panel shows this for a heavy-light (strange) system. The time separation was chosen to be $\Delta T = 22a$ for the heavy-light and $\Delta T = 32 a$ for the heavy-strange system. We note that the scale on the y-axis is the same in both panels.}
  \label{fig:bag_fits}
\end{figure}

\begin{figure}
  \includegraphics[width=\textwidth]{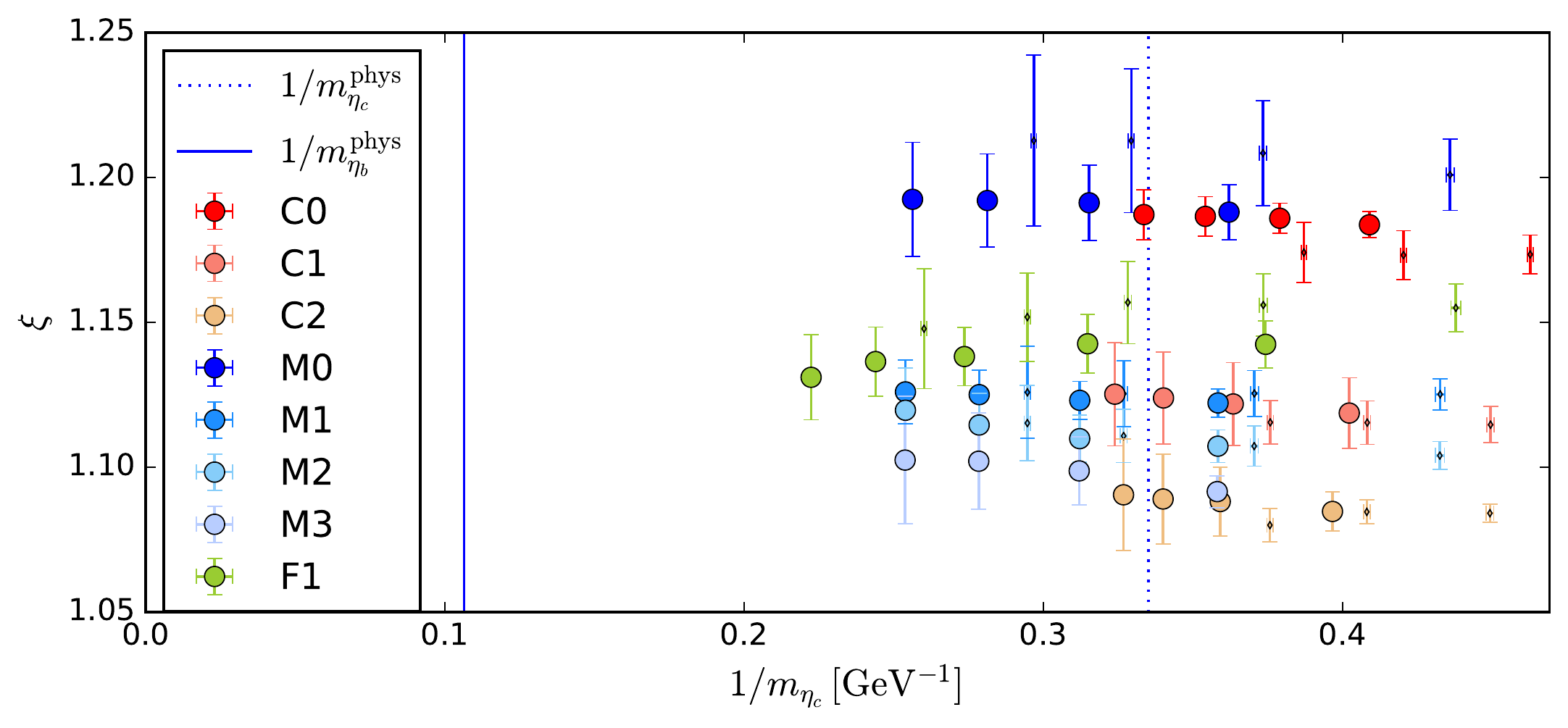}
  \caption{Collated fit results for the ratio $\xi$ on the different ensembles as a function of the inverse $\eta_c$ mass.}
  \label{fig:xi_dat}
\end{figure}

Similar to the correlation function fits presented in ref \cite{Boyle:2015kyy}, we build ratios $R(t,\Delta T)$ defined by
\eq{
  R(t,\Delta T) = \frac{\avg{\bar{P}_q^0(\Delta T) O_{VV+AA}(t)P_q^0(0)}}{8/3 \avg{\bar{P}_q^0(\Delta T-t)A_0(\Delta T)}\avg{A_0(t) P_q^0(0)}} \stackrel{0 \ll t \ll \Delta T \to \infty}{\longrightarrow} B_P^\mathrm{bare}.
}
Figure \ref{fig:bag_fits} shows example fits to the folded data $R(t,\Delta T)$ for the heaviest mass point on M0 for a light-heavy (left) and a strange-heavy (right) system. We investigated a number of different choices of $\Delta T$ and the chosen fits are representative choices ensuring that no excited state contamination is observed. We note that the y-scale is the same for the two panels, indicating that (given equal statistics unsurprisingly) the error on the heavy-light system is significantly larger than for the heavy-strange one. In the absence of the required renormalisation constants we investigate the ratio $\xi$ that is of phenomenological interest. The results are shown in Figure \ref{fig:xi_dat}. We again observe a very mild dependence on the heavy quark mass. We note that the error on $\xi$ is dominated by the error of the ratio of decay constants. Finally, we note that on the two physical pion mass ensembles (C0 and M0) $\xi$ is in good agreement, indicating mild continuum limit scaling.

\subsection{Charm quark mass}\label{subsec:charmmass}
Finally, we report on our efforts to determine the charm quark mass from our data.
Obtaining the renormalisation constant $Z_m$ for the choice of action of the \emph{Run 1} data is still work in progress.
However, the change in action between the light and the heavy discretisation is only minor and (at least in the case of $Z_A$) deviations between the light action and the mixed action have been at the sub-percent level~\cite{Boyle:2017jwu}. So for the purpose of the preliminary results presented here, we instead use the renormalisation constants of the light action which can be found in ref. \cite{Blum:2014tka}.

All results presented are preliminary and contain statistical errors only. For this reason we do not convert to the $\overline{MS}$ scheme and do not quote a final value. For illustrational purposes we only present the analysis in the $\slashed q$ scheme, but the qualitative features presented here are the same in the $\slashed \gamma$ scheme. Furthermore, the results from both schemes produce compatible results in the continuum limit. A detailed analysis including different hadronic input to fix the charm quark mass and a full systematic error budget will be carried out as soon as the renormalisation constants become available.

\begin{figure}
  \center
  \includegraphics[width=.45\textwidth]{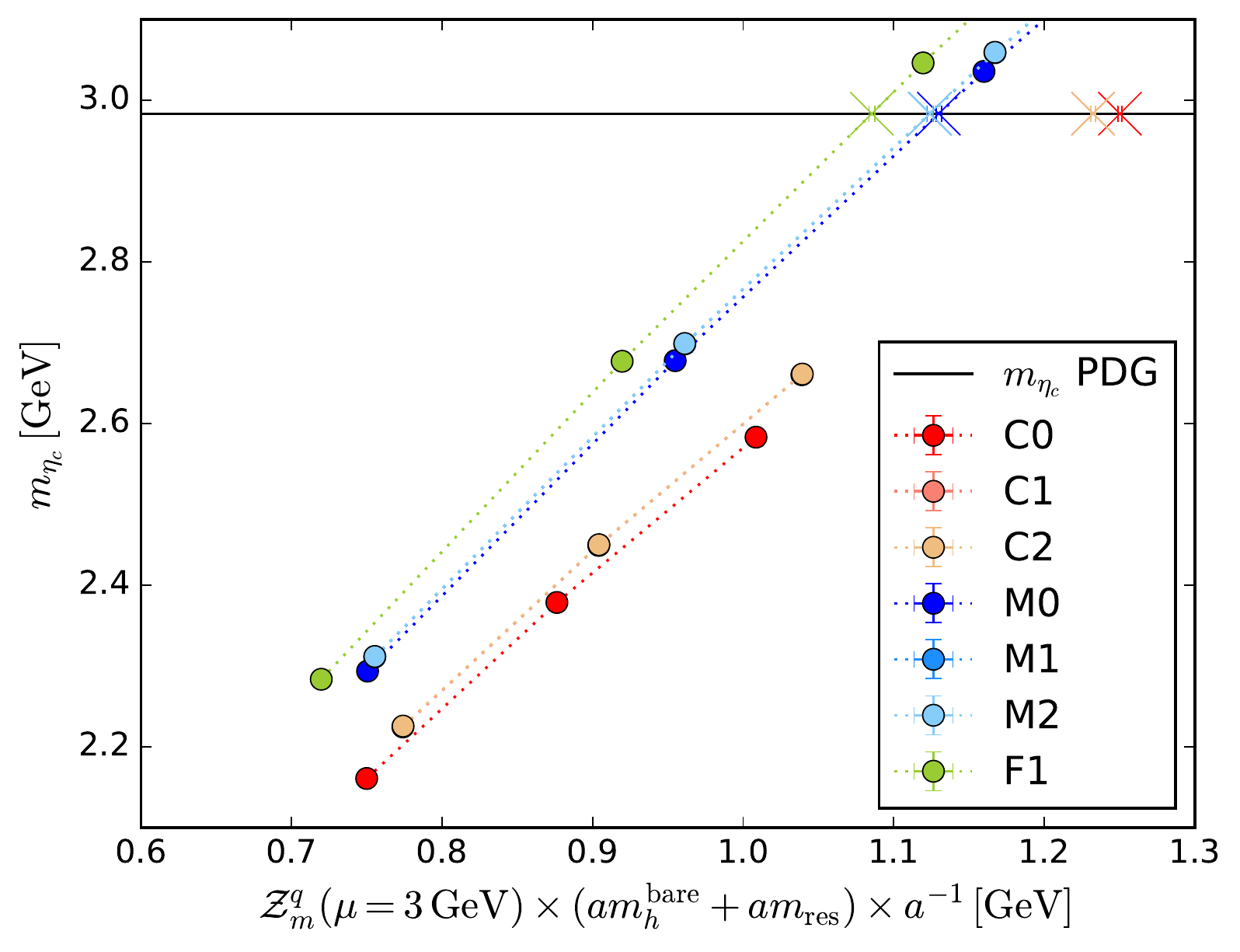}
  \includegraphics[width=.45\textwidth]{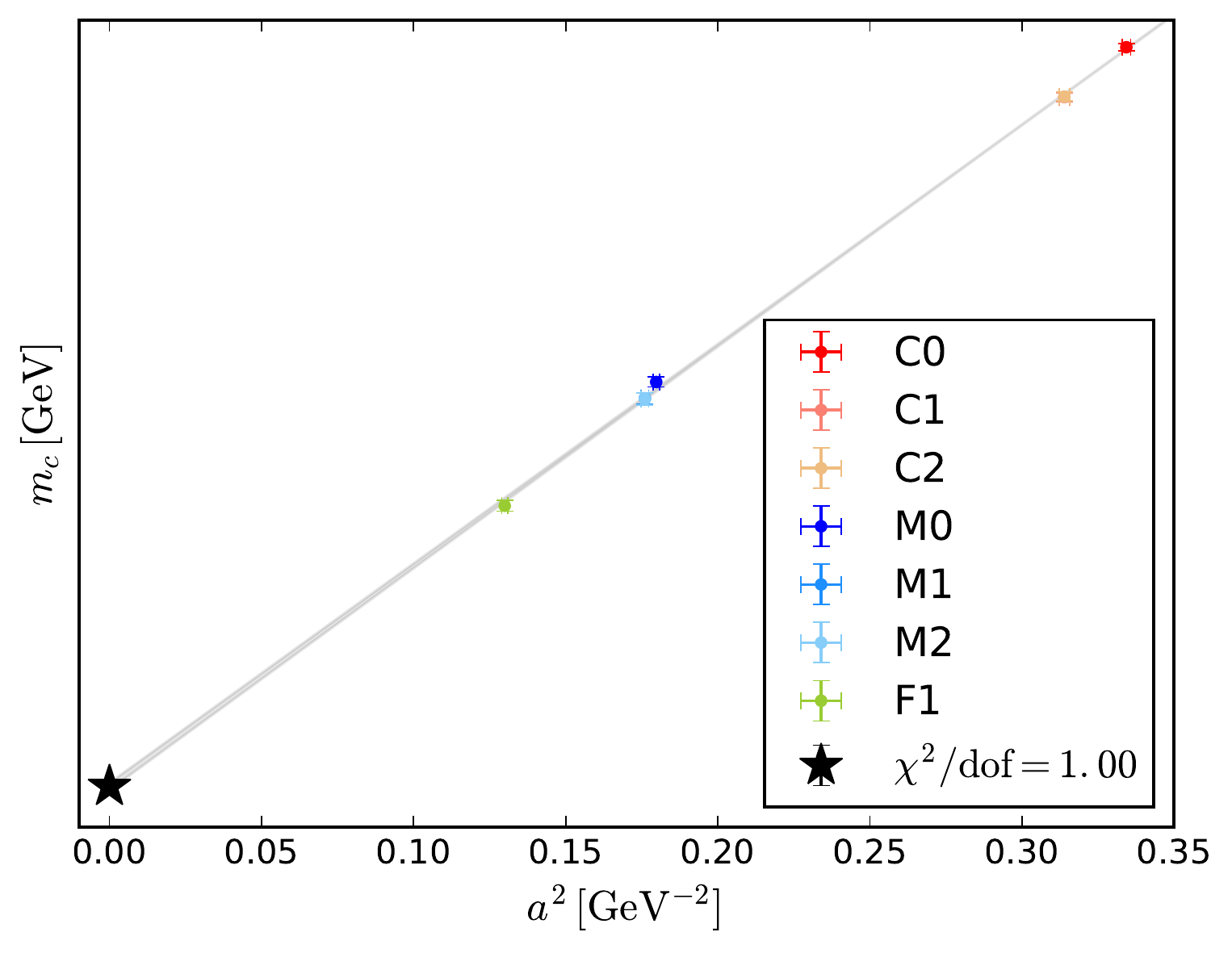}  
  \caption{\emph{Left}: $\eta_c$ mass as a function of the \emph{preliminarily renormalised} charm quark mass in the $\slashed q$ scheme. $m_{\eta_c}$ is read off from the intercept with the physical $\eta_c$ mass and indicated by the coloured crosses corresponding to the various ensembles. The physical $\eta_c$ value is given by the black solid horizontal line. \emph{Right}: Continuum limit of the renormalised quark mass as determined from the $m_{\eta_c}$ mass.}
  \label{fig:charm_mass}
\end{figure}

The left panel of Figure \ref{fig:charm_mass} shows the simulated values of the $\eta_c$ mass closest to its physical value as a function of the renormalised heavy quark mass. The horizontal black solid line shows the PDG~\cite{Patrignani:2016xqp} value to which the simulated data is inter/extrapolated. For the medium and fine ensembles, we simply linearly interpolate to the physical $\eta_c$ mass. On the coarse ensembles we do a linear fit to all three heavy mass points. The results of these mass inter/extrapolation are shown by the large crosses along the line indicating the physical $m_{\eta_c}$ mass. In a second step, we carry out a continuum limit fit to the data as shown in the right panel of the same figure. On the coarsest ensemble we observe discretisation effects of the order 25\%, however a continuum limit ansatz which is linear in $a^2$ describes our data well.

We are planning to supplement this analysis also with the \emph{Run 2} data as soon as the respective renormalisation constants become available. This will remove the need to extrapolate in the charm quark mass on the coarse ensembles. It is also expected that the discretisation errors from the second dataset will differ, allowing for two separate continuum limits with a universality constraint that will allow to test the validity of the described continuum limit ansatz.

\section{Summary and prospects}
We report on the status of RBC/UKQCD's domain wall charm physics program including three lattice spacings and two physical pion mass ensembles. In addition to the previously presented data set we have obtained a complete second data set with a different choice of heavy quark discretisation. This allows to reach the physical charm quark mass even on the coarsest ensemble. Furthermore, on the finest ensemble, we are now able to obtain heavy-heavy pseudoscalar masses extending up to roughly half the physical $\eta_b$ mass. This observation, combined with the mild behaviour with the heavy quark mass, raises hopes that we will be able to extrapolate our data to the physical bottom mass regime, providing $b$-physics predictions from domain wall fermions.

In particular we have reported on the status of the calculation of heavy-light and heavy-strange decay constants, neutral meson mixing and the charm quark mass from heavy quarks with improved statistical errors compared to our previous results. In parallel we also investigate the connected charm contribution to the Hadronic Vacuum Polarisation. Finally, we are also working on an analysis combining our domain wall fermion result with the data set produced by the JLQCD collaboration which has access to finer lattice spacings, but only at unphysically heavy pion masses. 

\begin{acknowledgement}
  The research leading to these results has received funding from the European Research Council under the European Union's Seventh Framework Programme (FP7/2007-2013) / ERC Grant agreement 279757,  the Marie Sk{\l}odowska-Curie grant agreement No 659322, the SUPA student prize scheme, Edinburgh Global Research Scholarship and STFC, grants ST/M006530/1, ST/L000458/1, ST/K005790/1, and ST/K005804/1, ST/L000458/1, and the Royal Society, Wolfson Research Merit Award, grants WM140078 and WM160035 and the Alan Turing Institute. The authors gratefully acknowledge computing time granted through the STFC funded DiRAC facility (grants ST/K005790/1, ST/K005804/1, ST/K000411/1, ST/H008845/1). 
\end{acknowledgement}

\bibliography{Lattice2017_209_TSANG}

\begin{thebibliography}{21}

\bibitem{Kobayashi:1973fv}
M.~Kobayashi, T.~Maskawa, Prog. Theor. Phys. \textbf{49}, 652 (1973)

\bibitem{Inami:1980fz}
T.~Inami, C.S. Lim, Prog. Theor. Phys. \textbf{65}, 297 (1981), [Erratum: Prog.
  Theor. Phys.65,1772(1981)]

\bibitem{Buras:1990fn}
A.J. Buras, M.~Jamin, P.H. Weisz, Nucl. Phys. \textbf{B347}, 491 (1990)

\bibitem{Allton:2007hx}
C.~Allton et~al. (RBC, UKQCD), Phys. Rev. \textbf{D76}, 014504 (2007),
  \texttt{hep-lat/0701013}

\bibitem{Allton:2008pn}
C.~Allton et~al. (RBC-UKQCD), Phys. Rev. \textbf{D78}, 114509 (2008),
  \texttt{0804.0473}

\bibitem{Blum:2014tka}
T.~Blum et~al. (RBC, UKQCD), Phys. Rev. \textbf{D93}, 074505 (2016),
  \texttt{1411.7017}

\bibitem{Boyle:2017jwu}
P.A. Boyle, L.~Del~Debbio, A.~J{\"u}ttner, A.~Khamseh, F.~Sanfilippo, J.T.
  Tsang (2017), \texttt{1701.02644}

\bibitem{Iwasaki:2011np}
Y.~Iwasaki (1983), \texttt{1111.7054}

\bibitem{Kaplan:1992bt}
D.B. Kaplan, Phys. Lett. \textbf{B288}, 342 (1992), \texttt{hep-lat/9206013}

\bibitem{Shamir:1993zy}
Y.~Shamir, Nucl. Phys. \textbf{B406}, 90 (1993), \texttt{hep-lat/9303005}

\bibitem{Brower:2004xi}
R.C. Brower, H.~Neff, K.~Orginos, Nucl. Phys. Proc. Suppl. \textbf{140}, 686
  (2005), [,686(2004)], \texttt{hep-lat/0409118}

\bibitem{Cho:2015ffa}
Y.G. Cho, S.~Hashimoto, A.~J{\"u}ttner, T.~Kaneko, M.~Marinkovic, J.I. Noaki,
  J.T. Tsang, JHEP \textbf{05}, 072 (2015), \texttt{1504.01630}

\bibitem{Boyle:2016imm}
P.~Boyle, A.~J{\"u}ttner, M.K. Marinkovic, F.~Sanfilippo, M.~Spraggs, J.T.
  Tsang, JHEP \textbf{04}, 037 (2016), \texttt{1602.04118}

\bibitem{Morningstar:2003gk}
C.~Morningstar, M.J. Peardon, Phys. Rev. \textbf{D69}, 054501 (2004),
  \texttt{hep-lat/0311018}

\bibitem{Tsang:2016iky}
J.T. Tsang, P.A. Boyle, L.~Del~Debbio, A.~Jüttner, A.~Khamseh, F.~Sanfilippo,
  O.~Witzel (RBC, UKQCD), PoS \textbf{LATTICE2016}, 278 (2016),
  \texttt{1611.06804}

\bibitem{Boyle:2015kyy}
P.~Boyle, L.~Del~Debbio, A.~Khamseh, A.~Jüttner, F.~Sanfilippo, J.T. Tsang,
  PoS \textbf{LATTICE2015}, 336 (2016), \texttt{1511.09328}

\bibitem{Boyle:2008rh}
P.A. Boyle, A.~Juttner, C.~Kelly, R.D. Kenway, JHEP \textbf{08}, 086 (2008),
  \texttt{0804.1501}

\bibitem{Martinelli:1994ty}
G.~Martinelli, C.~Pittori, C.T. Sachrajda, M.~Testa, A.~Vladikas, Nucl. Phys.
  \textbf{B445}, 81 (1995), \texttt{hep-lat/9411010}

\bibitem{Sturm:2009kb}
C.~Sturm, Y.~Aoki, N.H. Christ, T.~Izubuchi, C.T.C. Sachrajda, A.~Soni, Phys.
  Rev. \textbf{D80}, 014501 (2009), \texttt{0901.2599}

\bibitem{Aoki:2010pe}
Y.~Aoki et~al., Phys. Rev. \textbf{D84}, 014503 (2011), \texttt{1012.4178}

\bibitem{Patrignani:2016xqp}
C.~Patrignani et~al. (Particle Data Group), Chin. Phys. \textbf{C40}, 100001
  (2016)

\end{thebibliography}

\end{document}